\documentclass{acm_proc_article-sp}
\usepackage{amsmath}
\usepackage{usenix,epsfig,endnotes}
\usepackage[colorinlistoftodos,prependcaption,textsize=tiny]{todonotes}
\usepackage{hyperref}

\usepackage{algorithm}
\usepackage{mathtools}
\usepackage{url}
\usepackage[noend]{algpseudocode}
\begin{document}

\date{}

\title{Oops!...I think I scanned a malware}


\author{
{\rm Ben Nassi}\\Dept. of Software and\\ Information Systems Eng.,\\Ben-Gurion University\\ of the Negev,\\Be'er-Sheva, Israel\\nassib@post.bgu.ac.il
\and
{\rm Adi Shamir}\\Computer Science \\department,\\Weizmann Institute\\ of Science,\\Rehovot, Israel\\adi.shamir@weizmann.ac.il
\and
{\rm Yuval Elovici}\\Dept. of Software and\\ Information Systems Eng.,\\Ben-Gurion University\\ of the Negev,\\Be'er-Sheva, Israel\\elovici@bgu.ac.il
} 
\maketitle

\thispagestyle{empty}
\subsection*{Abstract}
This article presents a proof-of-concept illustrating the feasibility of creating a covert channel between a C\&C server and a malware installed in an organization by exploiting an organization's scanner and using it as a means of interaction. We take advantage of the light sensitivity of a flatbed scanner, using a light source to infiltrate data to an organization. We present an implementation of the method for different purposes (even to trigger a ransomware attack) in various experimental setups using: (1) a laser connected to a stand (2) a laser carried by a drone, and (3) a hijacked smart bulb within the targeted organization from a passing car. In our experiments we were able to infiltrate data using different types of light sources (including infrared light), from a distance of up to 900 meters away from the scanner. We discuss potential counter measures to prevent the attack.

\section{Introduction}
The popularity of computer scanners has increased significantly since they were first introduced, and today they are commonly used by organizations and individuals to scan images, handwriting, as well as printed text to a digital image \cite{ScannerHistory}. Considered the successors of traditional fax input devices, scanners are based on the concept of telephotography, however instead of enabling the transmission of simple text, entire images can be transmitted by a scanner. In recent years, different types of scanners \cite{ImageScanner} have been introduced, including: sheet-fed scanners, integrated scanners, drum scanners and even portable scanners. Probably the most popular type of scanner is the flatbed scanner, sometimes called a reflective scanner, which works by shining white light onto the scanned object and reading the intensity and color of the light that is reflected from it. Even today, nearly three decades since they were initially introduced, the popularity of flatbed scanners continues, and they are sold as an integrated part of a multi-function printer and as a standalone device. Even in the new era of smartphones, scanners continue to exist because they provide higher resolution output than any other device. 
\par In this article we present a novel method to infiltrate data into an organization. Our method uses light transmitted by an attacker to a flatbed scanner, which is then extracted by a malware installed in the organization. Our method exploits an organization's scanner which serves as a gateway to the organization, in order to establish a covert channel between a malware and an attacker. The attacker controlling the light source can be located far away from the targeted scanner. We present different experimental setups using several types of light sources and varying visibility and distance from the targeted scanner: (1) an  external laser installed on a stand positioned far away but within clear sight of the scanner, (2) an external laser connected to a micro-controller installed on a drone, remotely controlled by an attacker to approach the scanner when clear sight to the targeted scanner is available at a very close distance and even by (3) an internal smart bulb within the organization that was hijacked by an Android device located in a passing car. 
\par The motivation to establish a covert channel for data infiltration can vary from the aim of triggering a malware to encrypt critical files (ransomware), a day before an important presentation, to harming an organization by deleting important files.
\par A persistent attacker eventually finds a way to bypass security measures in order to infect a computer in the organization with a malware. Even isolated networks are not protected against malware infection as was shown in the cases of Stuxnet\cite{Stuxnet} and the Equation Group APT\cite{EquationGroup}. The main challenge of the attacker is to find ways to control the malware without being detected. Given a malware/bot installed on organizations' internal network, the proliferation of flatbed scanners and their connection to the network of the organization, countless organizations are vulnerable to our attack; the equipment required to establish the covert channel can be purchased for less than \$20 on eBay.
\par The contribution of this research is our ability to exploit a legitimate flatbed scanner to establish a covert channel between a C\&C server and a malware (previously installed in the organization). This proposed covert channel can evade existing security measures that focus on monitoring suspicious cyber-attack activities using commercial tools such intrusion detection and prevention systems (IDS/IPS), firewalls, and data leakage prevention (DLP) systems.
\par The paper is structured as follows: section \ref{sec:relatedwork} provides an overview of the related work, and section \ref{sec:scanner} contains the necessary background to understand the scanning process and the vulnerability exploited by the covert channel. Section \ref{sec:attack} describes the general schema to establish the covert channel, the parties involved, the challenges, and the protocol. Sections \ref{sec:Attacker}, \ref{sec:APT} describe the algorithms of the attacker and the malware. Section \ref{sec:analysis} presents analysis of the factors involved in the attack. Sections \ref{sec:Infiltrating commands with a clear line of sight}\ref{sec:Infiltration of commands with no clear line of sight to the scanner} present different implementations of the the attack: (1) when a clear line of sight to the scanner is available, and (2) when a clear line of sight is not available. In section \ref{sec:counter measures} we discuss counter measures, and in section \ref{sec:summary} we provide a summary.

\section{Related Work}
\label{sec:relatedwork}
In this section we describe the topic of covert channels and air-gapped networks and present related work in this area. The term covert channel was coined by Lampson \cite{CovertChannels,a-note-on-the-confinement-problem} who defined it as creating a capability to transfer information between parties that are not supposed to be allowed to communicate by measures that were not designed for communication. Covert channels can be used to circumvent system and network policies, by establishing communication that has not been considered in the design of the computing system for the purpose of communicating under the radar.
\par There are many ways to establish a covert channel between two parties. In some cases a bidirectional channel is established in which two parties can communicate fully. Other methods include a one way channel whereby only one party can transmit data, and the second party can receive it (broadcasting). The unidirectional covert channels can be used for various purposes including: (1) exfiltration of data from an organization (2) infiltration of data to an infected computer inside the organization's network from a C\&C server.
\par One well-known method of establishing a covert channel is by exploiting acoustical emanations. Previous studies \cite{o2014bridging,DBLP:journals/jcm/HanspachG13,185183} presented a method to establish a  bidirectional covert channel between two parties using inaudible sound. The main disadvantage of this method is that sound deteriorates with distance, so the parties must remain a reasonable distance from one another.
\par Another famous method used to establish a covert channel exploits optical emanations. A research over a decade ago, \cite{Loughry:2002:ILO:545186.545189} explored information leakage using devices' LED lights. They classified devices (such as modems, storage devices, CD-ROM drives and others) into three classes based on how informative the device's LED projection is (in term of information that can be gathered by watching it). More recent work \cite{7467343}, showed how to establish a unidirectional channel to exfiltrate data from an organization using a smart bulb. Other research used the same principle to exfiltrate data using a monitor's LED indicator \cite{Sepetnitsky:2014:EIA:2706698.2707281}.
\begin{figure*}
\centering
\includegraphics[width=1.0\textwidth]{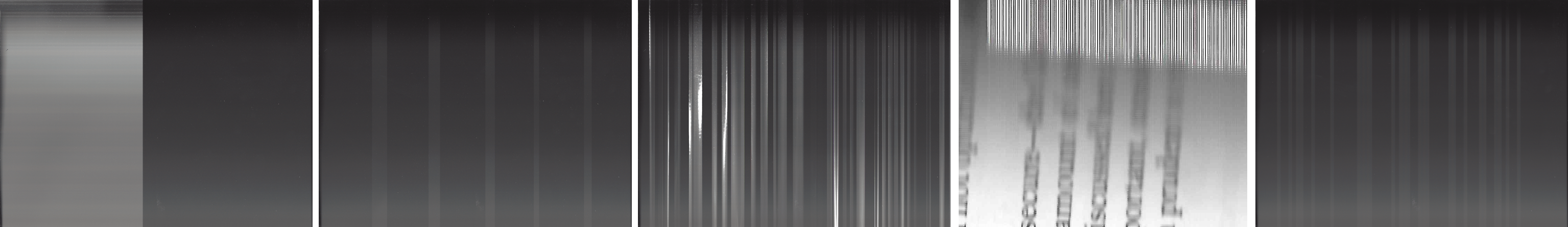}
\caption{Sensitivity of scanners to different light sources (from left to right): (1a) alternating room's light, (1b) alternating flashlight, (1c) alternating infrared laser pen, (1d) alternating laser on a document (1e) alternating screen saver of a monitor}
\label{fig:initial}
\end{figure*}
\par Less well-known means of establishing a covert channel used temperature as an indicator of the bits. Recent research \cite{190938} showed how to establish a thermal channel between different cores in multicore platforms to create a bidirectional channel. Other research \cite{7243739} used the same principle of thermal channels to exfiltrate data from one computer to another assuming they are positioned side by side. In both cases, the covert channel is based on a process that either heats or cools a device and a sensor to measure the change of temperature. This measurement is then demodulated to bits. This method primary disadvantage is its low transmission rate which is a result of the time needed to hit and cool the device.
\par Covert channels have also been investigated in air-gapped networks where they are used for communication with a compromised computer. Air-gapped networks are isolated networks that are not connected to the Internet and don't provide a way to communicate with computers that are outside of the air-gapped network. They are primarily used within closed organizations (e.g. military), in order to prevent confidential data from being transfered outside the network. In this context, covert channels are used in order to bridge the air gap and exfiltrate data from an infected computer connected to the air-gapped network to a device located outside the network, in the outside world. Another research \cite{guri2014airhopper} showed how to establish a unidirectional covert channel by transmitting a radio signal from a compromised isolated computer using the display video unit to a mobile phone with a radio receiver. In other work \cite{190936} researchers established a unidirectional covert channel by exfiltrating data from an isolated computer to a mobile phone using a software that transmitted electromagnetic signals at cellular frequencies by invoking specific memory-related instructions and utilizing the multi-channel memory architecture to amplify the transmission. 
\par In this article we present a novel and relatively easy to implement scheme to establish a covert channel that uses different light sources (such as laser and infrared beams) to interact with a malware that has been previously deployed in a network. In the rest of the paper we refer to our method in the context of bridging an air-gapped network, although it is not limited to this type of network, and it can be deployed in any other network.

\section{Sensitivity of Flatbed Scanners}
\label{sec:scanner}
In this section we describe how a flatbed scanner works \cite{FlatbedScannerWorks,FlatbedScannerWorks2,ScannerHistory,ImageScanner}. A flatbed scanner is made up of a lamp that is passed over a pane (from the bottom) to illuminate the scanner's pane. Using a series of lenses and mirrors, the light is bounced back to a light sensory array (CCD). A lens splits the image into three colors and the associated electrical charge is measured. The brighter the light reflected, the greater the electrical charge. Finally, an ADC device converts the electrical charge to a binary code that represents the document that is located on the pane. The scanning process results in a file in a configured format (such as PDF, PNG, etc.) that is transferred to a computer for storage by a direct physical connection between the scanner and the computer or using a network connection (wired or wireless).
\begin{figure}
\centering
\includegraphics[width=0.4\textwidth]{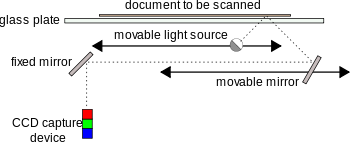}
\caption{Scanning process. Taken from Wikipedia}
\label{fig:scanner_process}
\end{figure}
\par Since the entire scanning process is influenced by the reflected light, interfering with the light that is illuminated on the pane will result in a different electrical charge which will therefore be parsed to a different binary representation of the scanned material. We wanted to examine the effect of external lighting on the scanner (during the scanning process). We conducted an experiment in which we used a flatbed scanner, which was partially open and located on a table. Figure \ref{fig:initial}a contains the output of a scan that was conducted in a room a in which the light was turned on at the beginning of the scan and turned off in the middle of the scan. Figure \ref{fig:initial}b shows an output of a scan that was taken when a flashlight of a mobile phone flickered on the pane. As can be seen, the scanner is sensitive to a change in the room's lighting, even the subtle change caused by the illumination of a mobile phone flashlight results in brighter shades. This light sensitivity causes the scanner to produce different shades in the output (binary representations) when the same experimental setup is used with different lighting conditions. The experiment was repeated with different flatbed scanners, and they were all found to be sensitive to external illumination even when a document was left on the their pane  (as can see in figure \ref{fig:initial}d. In addition, the scanners were sensitive to all of the kinds of lights used (i.e. the color of the light did not impact the results). Another  interesting observation is that scanner light sensitivity is not limited to the visible spectrum, but also includes infrared light. Figure \ref{fig:initial}c contains the output of a scan that was taken while an infrared laser pen was flickered on it. An infrared light, that can't be detected by the human eye causes the same effect on scanner output.
\par Since the output of the scan is highly influenced by external illumination, it can serve as the infrastructure to deliver messages from an external attacker to an organization that contains a flatbed scanner.
\section{The Attack}
\label{sec:attack}
\begin{figure*}
\centering
\includegraphics[width=0.8\textwidth]{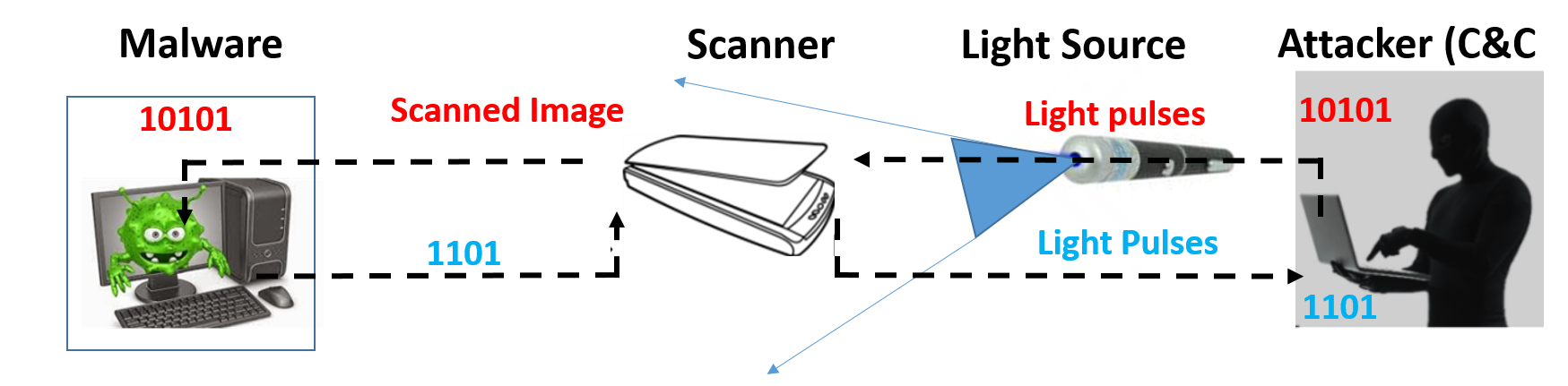}
\caption{Information Flow}
\label{fig:schema}
\end{figure*}
In this section we describe the threat model, the parties involved, the interactions, the assumptions, the challenges, and finally the considerations that led us to select the protocol to transmit data as a light sequence. 
\subsection{Threat Model}
\label{sec:threat_model}
Our attack's goal is to use the scanner as a gateway to an organization/home in order to infiltrate data/commands to a malware that has been deployed on a compromised computer/network. We use the scanner as a way to establish a covert channel between an external attacker and a malware installed in an organization. In this attack there are four parties involved:
\begin{enumerate}
\item Attacker's computer - This is a C\&C server operated by the attacker that controls a light source for the purpose of modulating commands. The attacker's control of the light source may be based on wired or wireless communication. 
\item Light Source  - The light source can be an external light source that is connected to a micro-controller and belongs to the attacker. The micro-controller modulates a given command (represented in binary code) from the C\&C server as a sequence of lights corresponding to the given command (the protocol used will be discussed later). The external light source and the micro-controller can be installed on a stand or even carried by a drone using a wireless connection to the C\&C (the attacker's PC).
Another option is to use an internal light source, located in the organization which the attacker has managed to control remotely (hijack). 
\item Flatbed Scanner - The flatbed scanner is located in the organization and is connected to the organization's network. 
\item Compromised Computer within the Organization -  Connected to the Organization's Network, This computer has been infected with a malware that can be used to command the scanner to lunch a scan and access the output (scanned digital image). The malware extracts the command sent from the C\&C and executes it.
\end{enumerate}
\par The attacker controls a light source that modulates the data as a “light signal” to a partially opened scanner during scanning. The scanning is launched by the malware in order to receive commands at known times (known both by the attacker and the malware). The malware extracts the command from the scan and executes it from the infected computer within the organization. The malware can then use the scanner to send an acknowledgment notification back to the attacker or to exfiltrate data from the organization by modulating the 1/0 bits as lights sent by launching scans.
\subsection{Assumptions}
\label{sec:assumptions}
We assume that the attacker has already successfully managed to infect a computer in an air-gapped network that has direct access to a scanner. Such a malware could be installed via a supply chain attack, by a malicious or unsuspecting insider, via active spear phishing emails, etc. Regarding the location of the scanner, we assume one of the following:
\begin{enumerate}
\item A clear line of sight from the outside to the scanner is exists - The scanner is located in a room with external wall. Two scenarios exist: (1) the wall  may allow outside light to pass (e.g., curtain wall), or (2) the wall contain a window.
\item A clear line of sight to the scanner is not exists - In this case, a device that produces light and can be hijacked is located in proximity of the scanner.
\end{enumerate}
It is important to note that we do not make any assumptions about the physical distance between the attacker and the organization (the two parties can be located far away from another or nearby). We also do not assume anything regarding the window's state (closed or open). 
\begin{figure*}
\includegraphics[width=0.7\textwidth]
{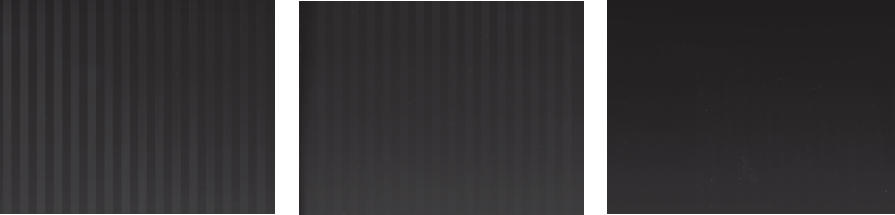}
\centering
\caption{ Output scans of the signal 1010..01 being transmitted by a laser from different distances (left to right): 3.2, 4.5, 7 meters.}
\label{fig:distances}
\end{figure*}
\subsection{Challenges and Solutions}
\begin{enumerate}
\item Physical Obstacles - The presence of obstacles can prevent an attacker from establishing a clear line of sight with the targeted scanner. For example this can occur when the targeted scanner is located on a very high floor that the attacker cannot reach or when the scanner is located in an isolated area of a building, far away from the attacker. Establishing a clear line of sight can be accomplished with the use of a drone equipped with the equipment required to conduct the attack and cellular connection that enables the attacker to send commands remotely; in addition, a powerful laser can be used to transmit the data in cases of long distance between the attacker and the targeted scanner.
\item Synchronization - Since the attacked network can be air-gapped, there may be no way to communicate with a malware in order to schedule scans. This problem can be solved by hardcoding the time of the scan to be fixed (e.g., every day at 23:00) in the code of the malware. Another option is to hardcode the time of first scan in the malware code and include the date of the next scan in the infiltrated data. The attacker can avoid minor time differences between the time that the scan was launched and the time that the attack launched using a camera to identify the beginning of the scan and trigger the attack. The camera may be installed on a drone or in another location within clear sight of the scanner.
\item Identifying a Scan - A problem can occur when an attack was not performed, and the malware must be able to identify these situations and ignore them. This problem can be solved by using a hardcoded prefix in each command as an identifier of the attack. 
\subsection{The Protocol}
\label{sec:protocol}
In order to establish a communication channel using the output of a scan a protocol that exploits a scanner's sensitivity to external illumination must be used. As was shown in section \ref{sec:scanner}, the effect of external illumination on the pane of the scanner results in brighter shades in the output of the scan (compared to the original shade). Given a command/data represented as a sequence of 1 and 0 (the original command is parsed to ASCI and represented in binary) and a light source pointed at the scanner or in proximity to the scanner, we modulate the bits as follows: we turn on the light source to indicate a 1 bit (i.e., illuminating on the pane) We turned off the light source to indicate a 0 bit. We pad the sequence with 1001 at both the beginning and the end of the command/data to be sent. Padding helps address potential problems of distinguishing between an empty packet and data that has not been transmitted). In addition, padding also helps identify the beginning and the end of the signal on from the output of the scanner and it allow the malware to calculate the window size of each bit of the scan (described in section \ref{sec:APT}).
\par Figure \ref{fig:distances} presents the original image that was scanned from the sequence of 1010...01 without padding using a laser that was controlled by a micro-controller. We found that different experimental setups, such as different distances, result in different shades to indicate a 1/0 bit. Adding padding that contains 1 and 0 bits helps the malware's code to identify the RGB value that associated with each of the bits. It allow the code to extract the values of the shades dynamically rather than making assumptions regarding them (described in section  \ref{sec:APT}).
\section{Attacker's Algorithm}
\label{sec:Attacker}
This section describes the attacker's algorithm. More precisely, we present how the algorithm that is deployed on the micro-controller (and connected to the light source) modulates a given command to attack a scanner according to the protocol discussed in subsection \ref{sec:protocol}.
\begin{algorithm}
\caption{Attacker's Algorithm}\label{attack_algorithm}
\begin{algorithmic}[1]
\Procedure{Attack}{command,window}
\State $\textit{cmd}  \gets \text{getInBinary}\textit{(commad)}$
\State $\textit{paddedCmd}  \gets \text{applyPadding}\textit{(cmd)}$ 
\State $index\gets 0$
\State $\textit{length}  \gets \text{length}\textit{(paddedCmd)}$
\While{$(index < length)$}
  \If{(paddedCmd[index] == 1)}{
    project() \;
  }
  \Else{
    dontProject() \;
  }
  \EndIf
\State $index\gets \text{index+1}$
\State $wait(window)$
\EndWhile\label{euclidendwhile}
\EndProcedure
\end{algorithmic}
\end{algorithm}
\par  Algorithm \ref{attack_algorithm} presents the stages of attack execution. The only parameter that is required to execute the protocol (besides the command itself) is a transmission window. First, the given command is represented as an array of 1/0 in binary (line 2). Then suffix and prefix padding are added to the binary representation as discussed in subsection \ref{sec:protocol} (line 3). Finally, the code is iterated over the array (lines 4-10). The light source is turned on to indicate a 1 bit and turned off the to indicate a 0 bit. The last command of the code is to wait a given period of time before proceeding to iterate. The reason for waiting before proceeding to the next iteration is to ensure that the light source is either opened or closed for the given amount of time that is denoted by the window (rate of a single bit). 
\par The rate of a single bit is primarily influenced by the ability of the scanner to absorb fast switches cleanly (describe in section \ref{sec:analysis}, the ability of the light source to switch between on/off states quickly and the speed of the scanning that adapts the required resolution. The speed of scanning can be measured by the attacker before the attack by performing a scan using the same type of scanner in the attacker's home/lab, or by reading the scanner's specifications. Given a scan speed and a sequence of bits (generated from a command), the window parameter can be computed by:
\begin{equation}
window = \frac{Time of scanning(ms)}{\#bits}
\label{eq:window}
\end{equation}
\makeatletter
\def\BState{\State\hskip-\ALG@thistlm}
\makeatother
\section{Malware's Algorithm}
\label{sec:APT}
This section describes the algorithm of the malware. As discussed in subsection \ref{sec:assumptions} the malware is previously installed on one of the organization's computer. The malware algorithm should be able to (1) execute a scan at a scheduled time and (2) extract a command from the scanned image. The algorithm's main challenge is to extract the command in various experimental setups, each of which will result in the same signal being scanned differently. As can be seen in figure \ref{fig:distances}, the white shades corresponding to the 1 bit deteriorates in each of the images, depending on the distance between the laser and the scanner, and therefore algorithm should be immune to shade's changes. Algorithm \ref{malware_algorithm} presents the signal extraction without hard-coded colors to identify 1 and 0 bits.
\begin{algorithm}
\caption{Malware's Algorithm}\label{malware_algorithm}
\begin{algorithmic}[1]
\Procedure{ScanAndExtractCommand}{()}
\State $\textit{path}  \gets \text{scan}\textit{()}$
\State $\textit{image [] []}  \gets \text{loadToRGB}\textit{(path)}$
\State $\textit{contrast [] []}  \gets \text{applyContrast(image)}$
\State $\textit{background }  \gets \text{getDominantColor(contrast)}$
\State $\textit{lineAverage []}  \gets \text{averageLines(contrast,background)}$
\State $\textit{threshold}  \gets \text{max(lineAverage)/2}$
\State $\textit{strechedSignal []}  \gets \text{strechSignal(lineAverage,threshold)}$
\State $\textit{paddedSignal []}  \gets \text{extractSignal(strechedSignal)}$
\State $\textit{signal []}  \gets \text{removePadding(paddedSignal)}$
\State $ \text{applyCommand(signal)}$
\EndProcedure
\end{algorithmic}
\end{algorithm}
\begin{figure*}
\centering
\includegraphics[width=0.9\textwidth]{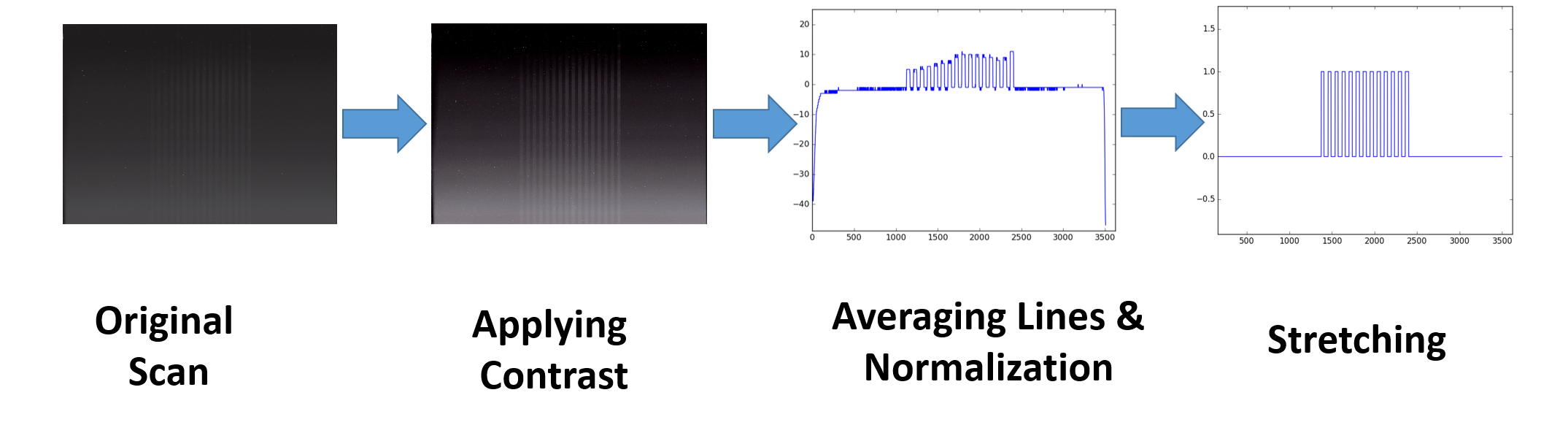}
\caption{Signal Extraction }
\label{fig:reconstruction}
\end{figure*}
This algorithm should be executed at a scheduled time that is known by the attacker (the attacker can hard-code the first scan date and supply the next date in each attack or hard-code all of dates alternatively). First, the algorithm triggers the scan (line 2) and loads the image to the memory as a two dimensional RGB array(line 3). A contrast function is applied to the image to emphasize the shade of the bits (line 4) and the background color (line 5) in is then extracted (the value with the most appearances in the matrix). A vector of lineAverage is computed (line 6); each index $i$ in the vector is calculated by averaging the line and normalizing the value according to the background value:
\begin{equation}
lineAverage[i] = \sum_{col=0}^{columns} (\frac{contrast[i][col]}{columns}) - background
\label{eq:lens}
\end{equation}
A stretching is applied to the values of lineAverage (lines 7-8) to produce a new signal, where each value in index i of the new signal is 1 (if it is larger than the threshold) or 0 (if it is smaller than the threshold). The signal is then extracted by identifying the padding prefix and suffix in the signal that starts and ends the signal, (line 9) and removing the padding from it(line 10). Finally, the command is executed (line 11).
\par The algorithm assumes that the infiltrated packet size is less than half the size of the image. It will ensure that the background will be selected as the 0 bit. The attacker should design the transmission rate of the bits such that it results in a scan that takes less than half of the scanning time (half of the length of the image). Different kind of algorithms can be used to extract the code. We chose this algorithm because it is simple to implement.
\par Figure \ref{fig:reconstruction} presents the signal extracted from the sequence of 101010...101. 
The positive effect of applying contrast to the image is demonstrated by comparing the image before the contrast to the image after contrast.
\section{Analysis of the Attack}
\label{sec:analysis}
This section presents an analysis of the four main factors in designing the attack: the type of light source to use, the distance between the scanner and the attacker, the rate of transmission, and the time of the attack.
\subsection{Influence of Light Source}
There are two types of light sources that can be used by the attacker to perform the attack: visible and invisible light. As mentioned in section \ref{sec:scanner}, scanners are sensitive to both types (see figure \ref{fig:initial}). In a case of an organization with a window with an IR filter (e.g., sunlight filters or a curtain wall), a visible light source can be used to attack the scanner. The selection of the specific light is primarily a function of the conditions of the room in which the targeted scanner is located. We prefer to use IR light whenever it is possible since it cannot be detected by the naked eye, making it an ideal covert communication channel.

\subsection{Influence of Distance}
\begin{figure}
\includegraphics[width=0.4\textwidth]{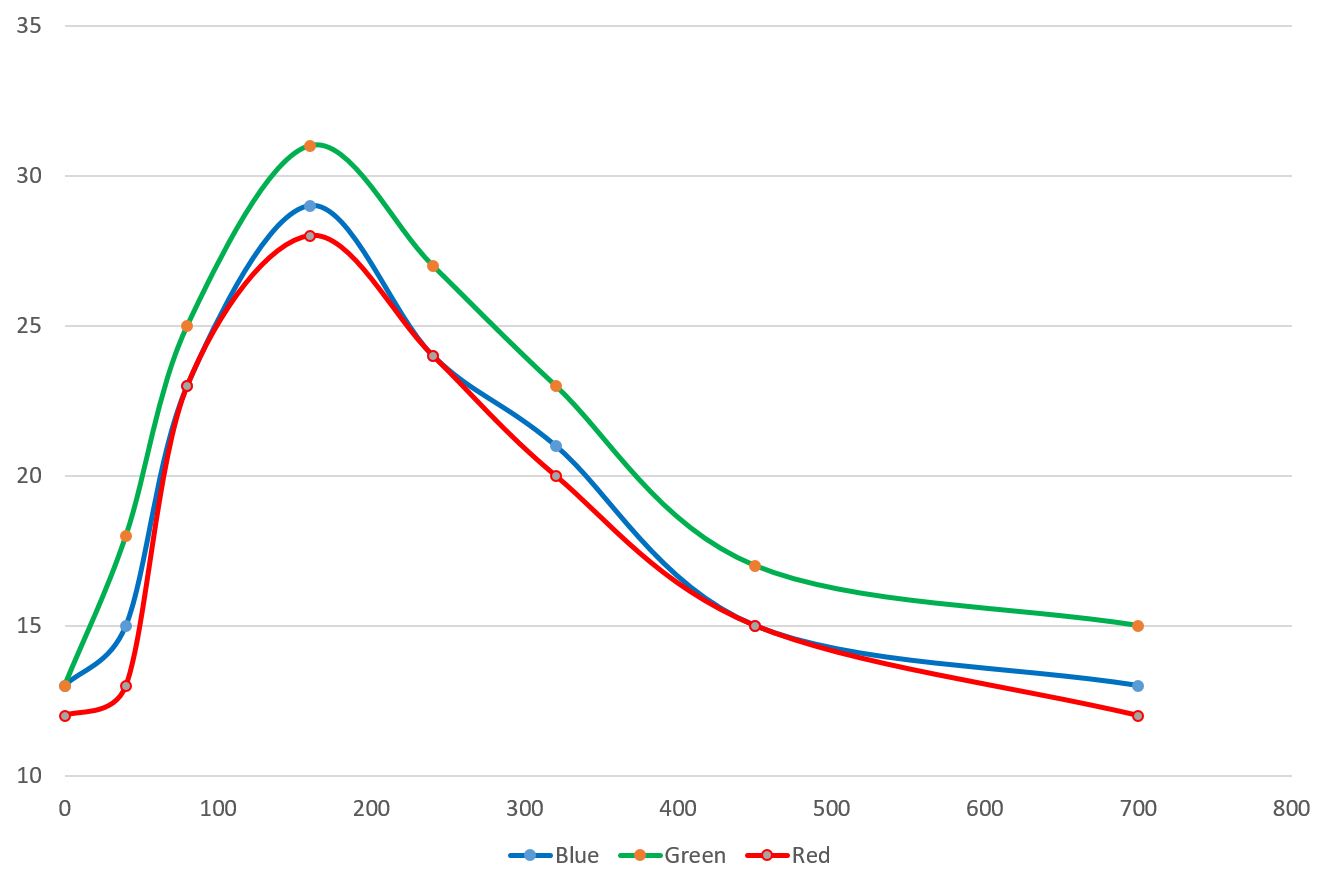}
\centering
\caption{Difference between 1 and 0 shades of Red, Green, Blue values from different distances. X axes represents the distance in cm between the attacker and scanner. Y axes represents the difference between 1/0 shades.}
\label{fig:parabola}
\end{figure}
Selecting the distance to attack a scanner using an external source light is highly influenced by the experimental setup (such as line of sight) and the intensity of the used light source. If the attacker has a clear line of sight to the scanner from his location he can use a stand to direct the source light (such as laser) to point to the pane of the scanner and perform the attack. Alternatively, the attacker can install the laser on a drone to minimize the distance in order to create a clear line of sight to the pane of the scanner and launch the attack remotely.
\par There are few issues to be considered. First, there is a trade-off between the stability that is required in order to attack a scanner from a distance of 10 meters away versus 100 meters away; the light source must remain stable in order to reach the scanner and perform the attack. This is extremely crucial when using a drone equipped with a light source to attack a scanner. 
\par Second, since light deteriorates with distance, the attacker need to use a light source that is strong enough to project the light the necessary distance. This need demonstrated in figure\ref{fig:distances} which shows how the quality of the scanned image is compromised, to the point that it cannot be extracted, when the distance between the attacker and the scanner increases.
\par Third, another important observation is that the distance of lighting influences also on the diameter of projection on the scanner pane. The projected diameter is decreases when the distance decreases. In our experiments we found that the best way to effect the result of a scan is with a diameter that covers the entire pane of the scanner. Since small distances between a laser and a scanner result in small diameters that is too focused in a specific dot on the pane it results in weak spreadness of the light on the pane. Again, the quality of the image is compromised, when the distance between the attacker and the scanner decreases under a threshold.
\begin{figure}
\includegraphics[width=0.4\textwidth]{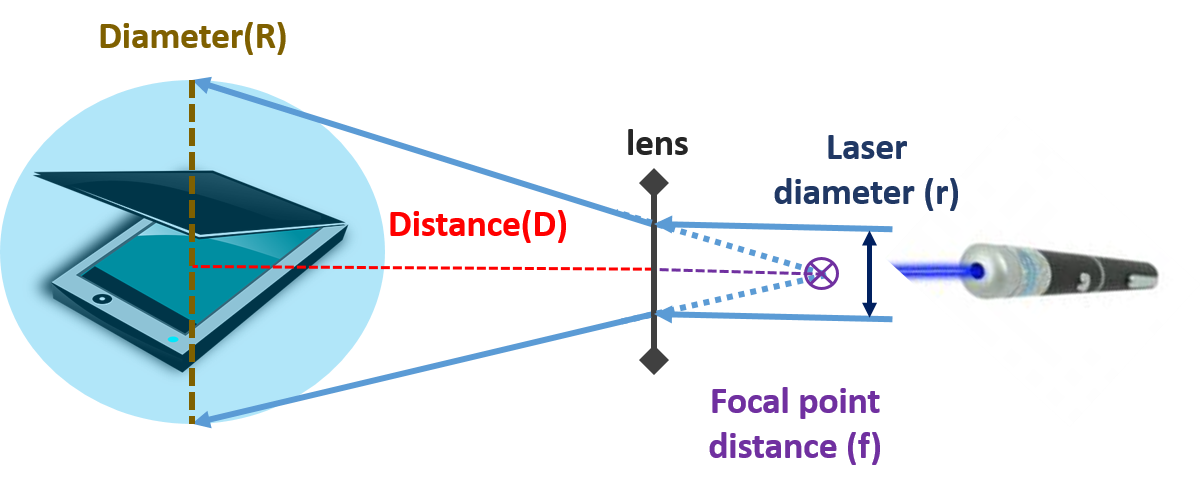}
\centering
\caption{Illustration of the focal (f), distance (D), diameter(R). The focal can calculated by equation \ref{eq:lens}}
\label{fig:diameter}
\end{figure}
\par The attacker's goal is to produce a clear scan (i.e maximize the differences between the 1/0 bits) so the malware will extract the signal correctly. The trade-off between the diameter of the projected light and the intensity of the light that is projected is presented on figure \ref{fig:parabola}. The figure presents the differences in the RGB values between the 1/0 shades as a function of the distance of projection. This graph was produced from seven experiments in which we used the same 300mW red laser under the same experimental setups from different distances (0-700cm). We extracted the shades of the 1/0 bits (from each output) and calculated the difference between them. The 1st interval of the graph, (0-40cm), shows a linear (even asymptote) behavior, as in this range the laser did not cover the scanner area. The 2nd interval (40-160cm) shows that increasing the distance increases the intensity of the shades as a result of a diameter optimization. The 3rd interval (160-450cm) shows that increasing the distance damaged the intensity of the laser. The 4th interval (450-700cm) shows linear (even asymptotic) behavior. The optimum distance to attack the scanner using the laser we used is from 160cm as it got the highest difference of RGB values between 1/0 bit. This optimum is achieved when achieving the minimal diameter to cover the pane. We computed the difference of the RGB shades as a function of the distance using Lagrange interpolation (equations \ref{eq:red},\ref{eq:green},\ref{eq:blue}.
\begin{equation}
\begin{multlined}
\label{eq:red}
Red =  -1.29^{15} x^7 + 2.61^{-12} x^6 - 2.03^{-09}x^5\\ + 7.88^{-07} x^4 - 0.0001 x^3 + 0.01 x^2 - 0.35 x + 12
\end{multlined}
\end{equation} 
\begin{equation}
\begin{multlined}
\label{eq:green}
Green =  1.07^{-18} x^7 + 9.08^{-14}x^6 - 1.65^{-10} x^5 \\+ 1.08^{-07}x^4 - 3.1^{-05}x^3 
 + 0.003x^2 + 0.03x + 13
\end{multlined}
\end{equation} 
\begin{equation}
\begin{multlined}
\label{eq:blue}
Blue = -1.37^{-15} x^7 + 2.7^{-12} x^6 -2.1^{-9}x^5 \\+ 8.36^-7 x^4 - 0.0001x^3 + 0.01 x^2 - 0.4x + 13
\end{multlined}
\end{equation} 
\par One option to focus the projection and maximize the intensity of the light is by using a lens on top of the light source. Figure \ref{fig:diameter} illustrates the required circle to cover the scanner and the parameters of distance and diameter. The required lens's focal point can be calculated as a function of the distance (D), the laser diameter (r) and the required projection circle (R) to cover the scanner using the next formula:
\begin{equation}
focal length= \frac{ r*D}{R-r} [meter]
\label{eq:lens}
\end{equation}
\end{enumerate}
\subsection{Influence of Transmission Rate}
As can be seen in figure \ref{fig:distances}, from a sequence of two alternating bits we can calculate the window (in terms of corresponding number of lines to a single bit) from the malware's perspective. When using padding, the first two bits are: 10, and the rate of transmission of a single bit can be calculated by identifying the first line of transmission of the 1 bit denoted by $first$, and the first line of transmission of the 0 bit is denoted by $last$ to calculate the window as follows:
\begin{equation}
Rate (\#lines) = first - last
\label{eq:rate}
\end{equation}
By calculating the rate dynamically we can change the rates of transmission in each scan in order to infiltrate more/less data (depending on the action we want the malware to apply) without the need to change the malware. The number of lines (transmission rate) that adapt to infiltrate a single bit are influenced by the speed of the scan and by the rate of transmission. We want to maximize the amount of the data that we are can infiltrate. The limitations from the attacker side on the maximum rate of transmission are influenced by the ability of the attacker's light source to switch its states between the on and off. From the malware side The limitation on the maximum rate of transmission is the ability to identify correctly the bits 1/0 bits to reconstruct the signal.
\begin{figure}
\includegraphics[width=0.4\textwidth]{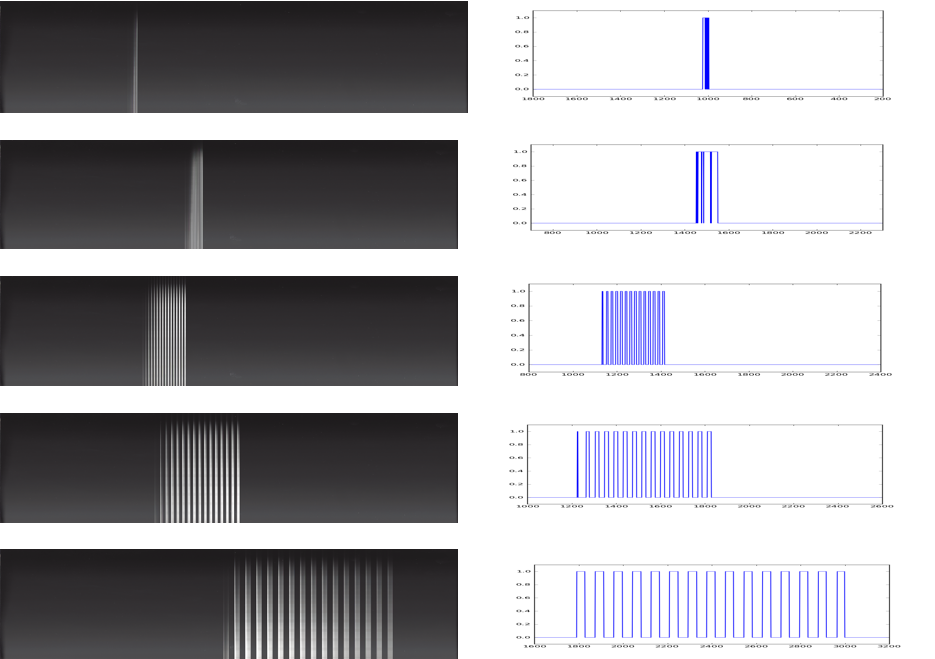}
\centering
\caption{Scans and Extracted signals in different rates (100ms,50ms,25ms,10ms,5ms)}
\label{fig:signals}
\end{figure}

\begin{table}[]
	\centering
	\resizebox{\columnwidth}{!}{%
		\begin{tabular}{ccc}
        \begin{tabular}[c]{@{}l@{}}			Transmission Rate   \end{tabular}

			& Extracted Signal
			& \begin{tabular}[c]{@{}l@{}}			Number of \\Errors \end{tabular}\\\hline
        	100 ms
			& 10101010101010101010101010101
			& 0\\\hline
            50 ms
			& 10101010101010101010101010101
			& 0\\\hline
            25 ms
			& 00101010101010101010101010101
			& 1\\\hline
            10 ms
			& 11111111111111111111111111111
			& 14\\\hline
            5 ms
			& 10011111111111111111111111111
			& 15\\\hline
		\end{tabular}
	}
\caption{Number of Errors of Extracting the signal 10101010101010101010101010101 in different rates of transmission.}~\label{tab:table2}
\end{table}
\begin{figure*}
\centering
\includegraphics[width=0.8\textwidth]{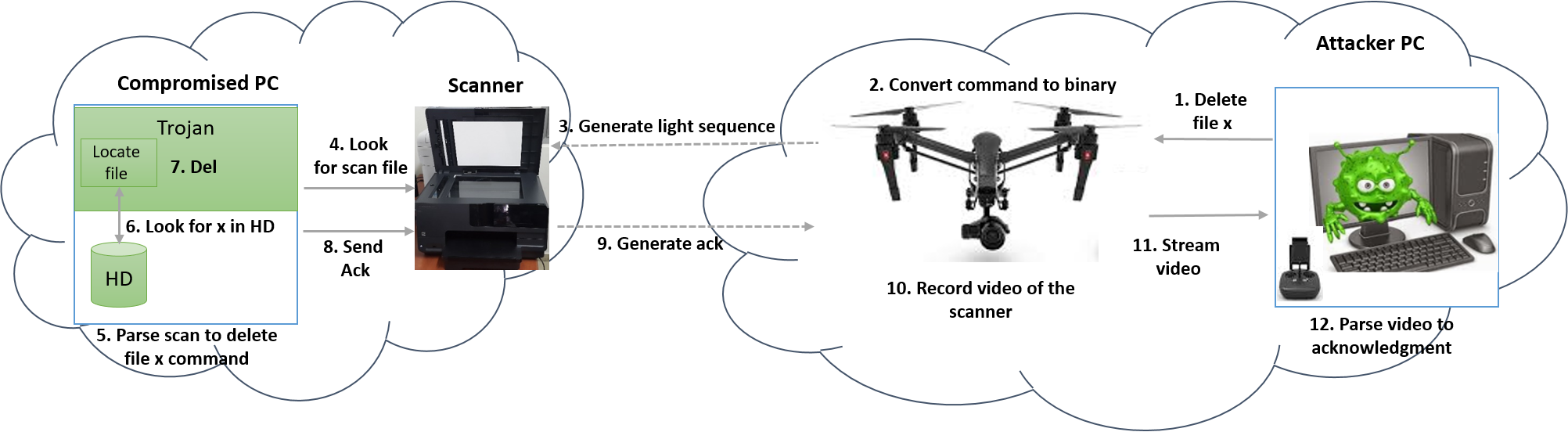}
\caption{Schema of the attack using external light source.}
\label{fig:schema}
\end{figure*}
\begin{figure*}
\centering
\includegraphics[width=0.8\textwidth]{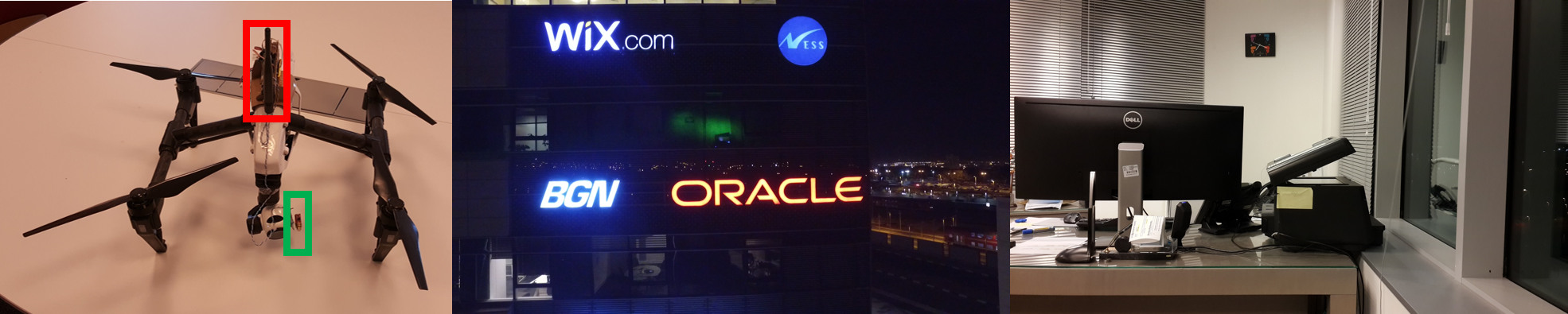}
\caption{Experimental setup (from left to right): (a) The Inspire 1 equipped with the laser (circled in green) and the Edison SoC, (b) the attacked organization during the attack, a green laser is projecting on a scanner behind a curtain wall in the 3rd floor, (c) the scanner in the office from the inside}
\label{fig:drone_experiment}
\end{figure*}
Figure \ref{fig:signals} presents the scanning and the extracted signal of the sequence 10101010101010101010101010101 with different transmission rates for a single bit (100ms,50ms,25ms,15ms,10ms,5ms). As can be seen from the figure, different transmission rates result in thinner sequence of columns that correspond to a single bit. 
Table \ref{tab:table2} presents the error (number of incorrect bits) of the extracted signal for each of the rates. Using our laser and our scanner, we found that maximum infiltration rate possible to use without any errors is 50 milliseconds. An error correcting code can be integrated to the malware's code to detect and correct the extracted code for rates faster than 50 milliseconds.
It is important to note that the speed of scanning can be changed by configuring different resolutions (Dots Per Inch) of the output scan. In the entire set of experiments described in this article we used standard quality of 300 dpi (default quality of the scanner), that lasted around 8 seconds. Higher resolution results in slower scanning and can help the attacker to infiltrate more data.
\subsection{Influence of Time}
We checked the influence of the hour of the day on the attack. We were able to perform the attack during day and night. Picking the hour to attack is only a consideration of minimizing the amount of people in the organization.
\section{Infiltration of commands with a clear line of sight to the scanner}
\label{sec:Infiltrating commands with a clear line of sight}
The first scenario demonstrates few attacks using an external light source when a clear line of sight to the scanner is available. We implement the first type of attack using an external light source connected to a micro-controller. The micro-controller can be installed on a stand if clear sight to the targeted scanner is available. Another possibility is installing the micro-controller on a drone to improve the sight of the target laser and control the micro-controller remotely using GSM communication from the C\&C. 
\par Figure \ref{fig:schema} presents the steps of the attack. An attacker's PC generates a sequence of binary signals transferred a micro-controller (phases 1-2).  The micro-controller launch the attack by modulating the received signal as light sequence using a connected laser that is directed to a scanner's pane while the scanner is scanning (phase 3). Next, the scan is extracted to a command (phases 4-5), and executed by the APT (phases 6-7). Finally, a message sent from the scanner to the attacker by modulating the bits of the message as scan and emitting visible light pulses from the flatbed (phase 8). The emitted light is captured by a drone equipped with a video recorder (maintaining a line-of-sight with the scanner). Finally, the video is processed by the attacker to extract the message from the malware.
\begin{figure*}
\centering
\includegraphics[width=0.8\textwidth]{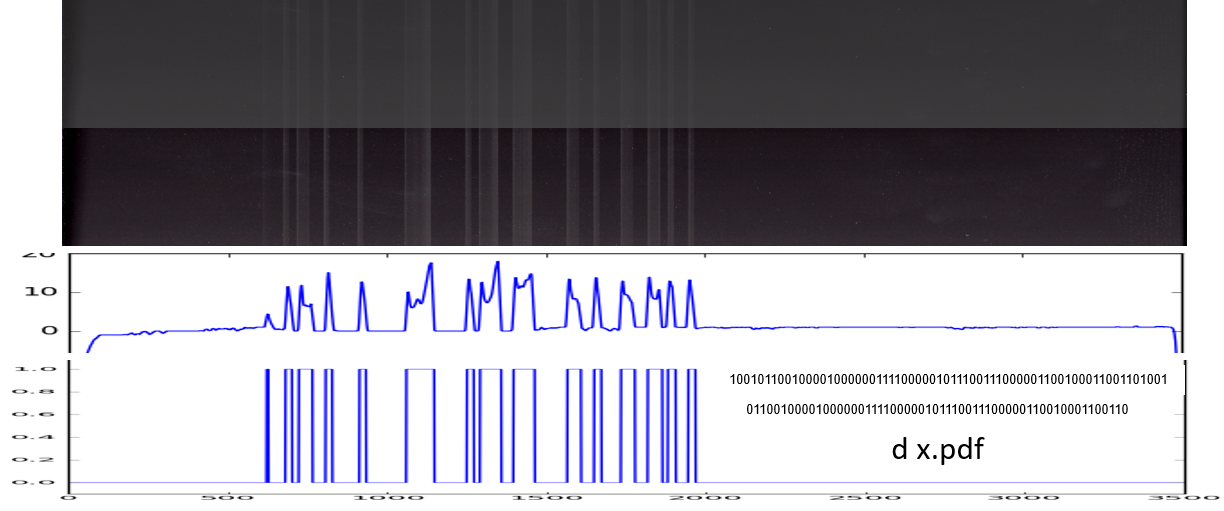}
\caption{Extracted Signal delete x.pdf, from top to bottom: (a) The original signal, (b) the signal after contrast, (c) line averaging (c) Stretching unpadding and ASCI representation}
\label{fig:d x}
\end{figure*}
\begin{figure*}
\centering
\includegraphics[width=0.8\textwidth]{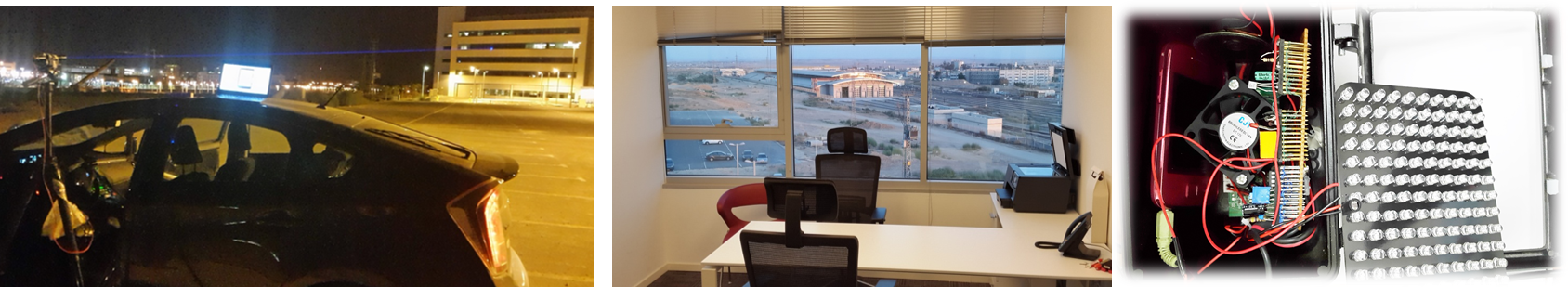}
\caption{Left to right: (a) The laser with the stand. (b) The attacked office with the scanner. (c) The IR projector.}
\label{fig:stand}
\end{figure*}
\subsection{Experimental Setup}
The described attack presents harming an organization, located in the 3rd floor in its building, by deleting an important file. The organization contained HP OfficeJet Pro8610 \cite{HP} a multi-function printer with a scanner and a fax connected to the network of the organization located on a table within a room as can be seen in figure (\ref{fig:drone_experiment})c. 
As can be seen in figure \ref{fig:drone_experiment}b, the building constructed from curtain walls that filters the sunlight and IR. So we decided to use a visible green laser that was dismantled from a laser pen (bought from eBay and operates on 600mW). We build an electrical circuit as follows: We took an Intel Edison \cite{Edison},  System on a chip (SoC), and connected it to a battery and the green laser. We installed the code described in section \ref{sec:Attacker} on the Edison. We installed the Edison SoC on top of the Inspire 1 pro (a drone) \cite{Inspire1}  while the laser was installed on its camera that was positioned on the stabilizer of the drone to maximize the stability during the attack. The drone equipped with the Edison and the laser can be seen in figure \ref{fig:drone_experiment}. The video from the camera of the drone was streamed to an Android application \cite{DJIGO} using radio signals. We established a communication channel between the C\&C computer and the Intel Edison SoC using its WiFi interface over a hot-spot that was opened from a mobile phone. The attacker had the ability to (1) launch the attack using the Edison SoC remotely from the C\&C and to (2) maneuver the drone using its joystick and watch the streamed video in an Android application.
\par As for the malware, we implemented the code described in section \ref{sec:APT} using a python  library Pyinsane \cite{Pyinsane}. We installed the malware on a computer inside the organization that was connected to its internal network. The malware scanned the network for a scanner (we found the IP of the scanner dynamically) and triggered by the scanner to launch a scan exactly at 23:00.
\subsection{Results}
The attack was recorded and uploaded to a link:\ref{DroneVideo}. We flied the drone to stand in front of the scanner in the 3rd floor from a distance of 15 meters and launched the attack after we detected that a scan was launched using the camera of the drone (23:00). We used a transmission rate of 50 milliseconds for each bit and infiltrated the command of "d x.pdf" where d is correlative to delete and x.pdf is the file to delete. Only 7 bytes are required (including the space) to encode this command. We padded the command with 1001 prefix and 1001 suffix (as described in section \ref{sec:protocol}. The padding added one more byte results in 8 bytes in total. The infiltration time took 3.2 seconds (64 bits that were transmitted in a rate of 50 milliseconds per bit). This was the entire time that the laser was used. The malware extracted the signal in real-time and sent an acknowledgment message by triggering another scan after the command was executed. Figure \ref{fig:d x} presents the output of each of  the stages of the signal extraction process as was described in section \ref{sec:APT}.
\footnote{\label{DroneVideo}\url{https://youtu.be/vy8dKaHNB-A}}
\par We repeated the experiment with much powerful laser (10W) that we installed on a dedicated stand from 900 meters from the targeted scanner (that was located in another room) as presented in figures \ref{fig:stand}a and \ref{fig:stand}b. We successfully infiltrated data to the scanner and received its acknowledgment using a telescopic camera. Finally, we repeated the same experiment with invisible light using an IR projector (the IR projector can be seen in figure \ref{fig:stand}c) from a distance of 20 meters and successfully infiltrated the command "erase file xxx.doc". As discussed in section \ref{sec:attack}, we could even increase the distance using a lens and a stronger laser.
\section{Infiltration of commands with no clear line of sight to the scanner}
\label{sec:Infiltration of commands with no clear line of sight to the scanner}
The second scenario demonstrates an attack using an internal light source (within the organization) when a clear line of sight to the scanner is not available and external light source can not be used.
We implement the second type of attack using an internal IoT device that produces light either directly or indirectly. More precisely, the attacker performs a functional attack (as was described in \cite{7467343}) on an IoT device inside an organization to produce the light sequence of a given command while the malware launches a scan and extracts the command from it.
\begin{figure*}
\centering
\includegraphics[width=0.8\textwidth]{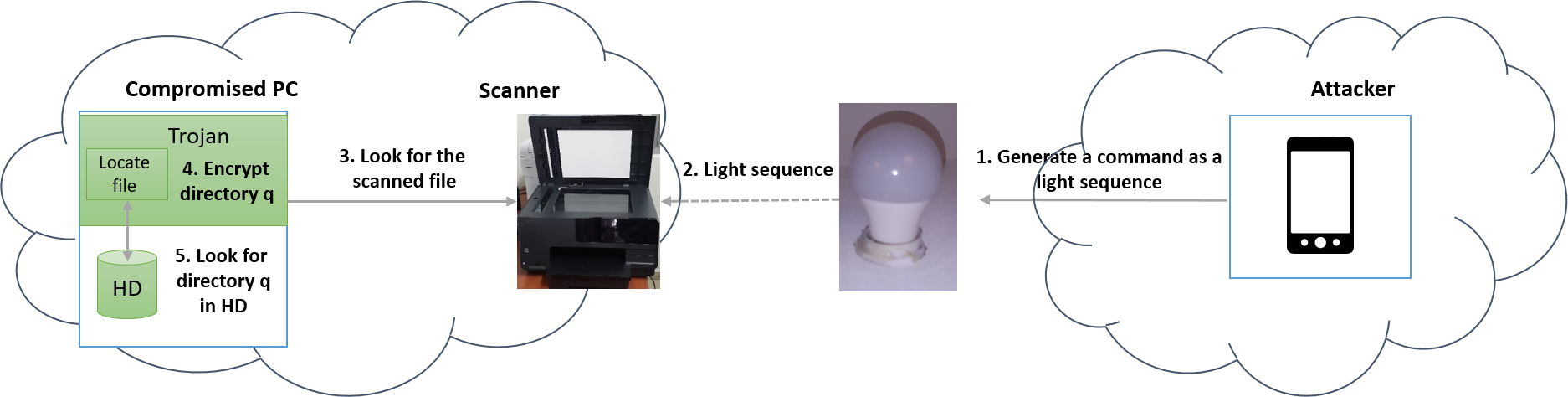}
\caption{Schema of the attack using hijacked internal light source}
\label{fig:iotSchema}
\end{figure*}
\begin{figure*}
\centering
\includegraphics[width=0.8\textwidth]{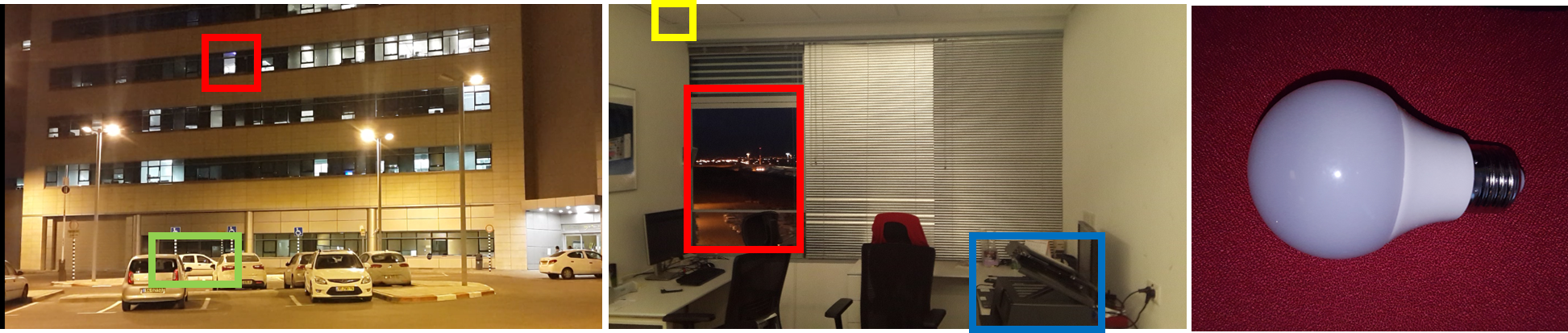}
\caption{Left to right: (a) Passing car (circled in green) attacks a scanner located in an office (circled in red), (b) The attacked office from the inside, the bulb (circled in yellow), the scanner (circled in blue) that is hidden behind a close curtain, (c) The hijacked smart bulb (MagicBlue)}
\label{fig:bulb}
\end{figure*}
\par IoT devices have become increasingly popular in the recent years and are widely sold. They can be controlled from various types of controllers (e.g., smartphones, PCs, smart remotes) using different protocols such as WiFi, infrared, Zigbee\cite{ZigBee}, Bluetooth, and Bluetooth Low Energy (BLE) \cite{BLE}. In practice, many commercial devices do not implement strong authentication mechanisms (some don't even implement any authentication mechanisms at all) exposing them to different kinds of attacks, including replay attacks (hijacking the device). For example, a television news report recently prompted viewers' Amazon Echo devices to order unwanted dollhouses \cite{AmazonAlexa}. The same thing happened to Google's assitant as a result of Google Home's Super Bowl ad\cite{GoogleHome}. Commercial Smart Bulbs were also found vulnerable to attacks on their bridge using external ZigBee signal \cite{DBLP:journals/iacr/RonenOSW16,DBLP:journals/corr/MorgnerMB16} and many BLE devices (even smart locks) were founded to be vulnerable to different kind of attacks, including hijacking\cite{BlackHatBLE,PickingBluetoothLowEnergyLocksfromaQuarterMileAway,Ho:2016:SLL:2897845.2897886}.
\par In this scenario the attacker exploits : (1) the unauthenticated protocols of IoT devices, and (2) the ability to produce a light from a specific IoT device which its light reflections can be received by the scanner. The attacker hijacks an IoT device with that produces lights in order to modulate the command to the scanner. The attacker can either attack an IoT device whose purpose is to illuminate (e.g., a smart bulb) or attack an IoT device in which illumination is a side effect, (e.g., a smart TV).
\par This attack scenario is preferable to the previously one described in the article, because there is no clear line of sight between the attacker and the scanner. However, in this scenario the attacker did manage to hijack an IoT device,located close enough to the scanner, in such a way that the produced lights from the IoT device effects the scanner's output. In this scenario the attacker attacks the scanner indirectly by attacking the IoT device. 
\par Figure \ref{fig:iotSchema} outlines the attack. First, the attacker hijacks an IoT device and use it to modulate the command as a sequence of lights. The scanner, which was scheduled to launch a scan by the malware, scans the reflections of the lights produced by the IoT device to an image and the malware extracts the signal from the scan and executes the command. In the current scenario the acknowledgment from the malware side is not part of the protocol (there is no clear line of sight to the scanner).
\subsection{Experimental Setup}
In our experiment we decided to attack a commercial smart bulb in order to infiltrate the required command to the organization via the scanner. Smart bulbs have been shown to be vulnerable to hijacking in various researches \cite{ReverseEngineeringBluetoothLightbulb,ReverseEngineeringBluetoothLightbulb2,ReverseEngineeringBluetoothLightbulb3,DBLP:journals/iacr/RonenOSW16}. The ability to control (hijack) a smart bulb inside an organization enables the attacker to exploit the known "on" and "off" commands to modulate the light sequence of a command.
\begin{figure*}
\includegraphics[width=0.8\textwidth]{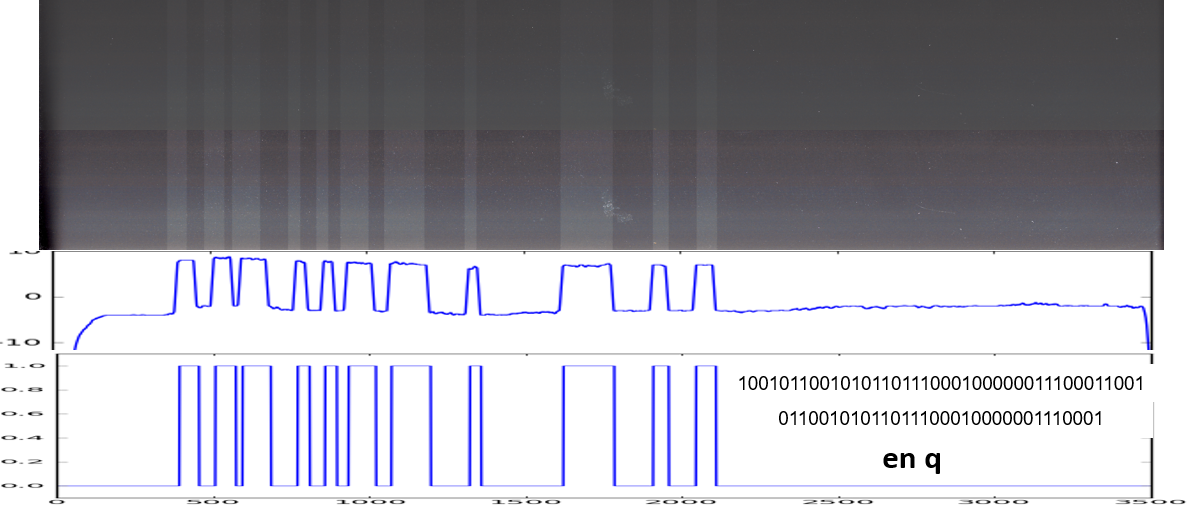}
\centering
\caption{Extracted Signal encrypt dir q, from top to bottom: (a) The original signal, (b) the signal after contrast, (c) line averaging (c) Stretching unpadding and ASCI representation}
\label{fig:en_q}
\end{figure*}
\par We attacked the same organization mentioned in the previous scenario. However, this time we attacked a different room which contained a scanner and a Magic Blue UU E27 Bulb\cite{MagicBlue} (presented in figure \ref{fig:bulb}c). The scanner, as can be seen in figure \ref{fig:bulb}b, is hidden behind a closed window and a clear line of sight does not exist. The MagicBlue smart bulb exposes BLE API for control. The BLE protocol has become quite popular for IoT devices, and it provides the ability to control from short distances (less than 100 meters). This bulb does not contain any type of authentication mechanism (as was found by \cite{ReverseEngineeringBluetoothLightbulb}). We reverse engineered by triggering commands using its official application \cite{MagicBlueApp} and sniffing the BLE traffic that was transmitted from the Android device. We used Wireshark \cite{Wireshark} to parse the PCAP file that was produced by the Android and found the commands:
\begin{center}
\begin{tabular}{ c c  }
 BLE Command & Functionality  \\\hline 
0xCC2333 & ON  \\\hline
0xCC2433 & OFF \\\hline
0x56FFFFFF00F0AA & Max Brightness \\\hline
0x5600000000F0AA & Min Brightness \\\hline
\end{tabular}
\end{center}
\par It is important to note that many kind of commercial smart bulbs are vulnerable to hijacking as was shown in the past \cite{ReverseEngineeringBluetoothLightbulb,DBLP:journals/iacr/RonenOSW16,ReverseEngineeringBluetoothLightbulb2,ReverseEngineeringBluetoothLightbulb3,DBLP:journals/corr/MorgnerMB16}. The contribution of this attack is that it shows that despite the fact that smart bulb does not contain any important information and might cause a minor damage if a external attacker has manage to hijack it (as opposed smart locks for example), the bulb can cause big damage when used as a mediator in attacks.
\par In our study we performed the attack in different experimental setups and different rates to produce the required light sequence in order to infiltrate the data. We found that even the slightest change of brightness (5\%) produced by the bulb can be detected by the scanner (e.g., changing the brightness of a bulb from 100\% to 95\%). Also, as described in section \ref{sec:analysis} we found that lights that produced by the bulb during a daylight were detected by the scanner. Also, we found that we are able switch the bulb states in frequencies less than 25 milliseconds, rates that will not be detected by the human eye, but as was shown in section \ref{sec:protocol}, we would have to add the malware's code an error correcting code for this purpose.
\subsection{Results}
Finally, we present the attack in the context of triggering a ransomware (with hardcoded key) given a directory, from a passing car (figure \ref{fig:bulb}a) on a scanner located in the 3rd floor (figure \ref{fig:bulb}b) behind a closed curtain with no line of sight using hijacked smart bulb, in the evening. The driver held a Samsung Galaxy S4 while driving in order to perform the attack from, a dedicated application that we wrote and installed on the Galaxy. The application scans for a MagicBlue smart bulb and connects to it. After connection, the application modulates a given command as light sequence using a series of "on" (1 bit) and "off" (0 bit) signals sent from over a BLE channel. The attack was recorded using a video camera and can be seen on the link\ref{BulbVideo}.
\footnote{\label{BulbVideo}\url{https://youtu.be/jHb9vOqviGA}} We used the same malware as described in \ref{sec:APT}, so the selected transmission rate was picked to be 100 milliseconds (that does not require any error correcting code). We infiltrated 40 bits (four bytes of data plus one byte of padding). The entire attack lasted for four seconds. The malware as explained in section \ref{sec:APT} is independent of the number of bits as well as the rate of transmission and does not require any error correcting code. The signal "en q" that is correlated to "encrypt directory q" was extracted by the malware in real-time and presented in figure \ref{fig:en_q}.
\section{Limitations}
Our method is effective with a partially open scanner or an open scanner. In our experiments we were unable to infiltrate a signal when the scanner was completely closed since light can not be projected on the pane when the scanner is closed. Since most of the organizations don't require closing scanners after each scan, and  it is not out the question that an organization's employees could get paid by an attacker to intentionally leave a scanner open, opportunities exist to use our method.
\section{Counter Measures}
\label{sec:counter measures}
Since manufacturers don't consider scanners a mean to infiltrate data, they offer Ethernet and WiFi connectivity and drivers in order to support remote and direct scanning. As was shown in section \ref{sec:evaluation}, limiting the line of sight to scanners by positioning them behind a wall doesn't provide a solution for the threat (shown in section \ref{sec:evaluation}).
\par Various solutions can be used to automatically prevent/detect the threat, rather than relying on the awareness of individuals within an organization to close the scanner after completing a scan including: (1) disconnecting the scanner from the network. Many scanners support scanning directly to a disk on key device using USB interface. Doing so, prevents an attacker from establishing the covert channel in this paper. This might be considered an extreme solution, since it also limits printing and faxing remotely, if the scanner is part of an all-in-one device. (2)  Extending the scanner's remote protocol to support two levels of authentication.
\par However, we believe that a proxy based solution will prevent the attacker from establishing such a covert channel without the need to apply extreme changes. The scanner will be connected by a wire directly (e.g., using a USB interface) to a computer (proxy) within the organization's network instead of being connected to the network. The proxy will provide an API. When a scanning request is received, the computer initiates a scan and processes the output in a classifier in order to detect malicious scan. If the scanning was classified as benign, the output is returned to the entity that requested the scan. Otherwise, the scan will be stored on the proxy for analysis of the IT team. The classifier's logic for detection of suspicious scans can be based on (1) robotic detection or anomaly detection by analyzing the characteristic of the scan (e.g. unusual hour of the day), or (2) raw data analysis (the scan output) using a machine learning model that was trained on malicious/benign scans.
\section{Summary}
\label{sec:summary}
In this research we showed how a simple organization's scanner can be used in order to establish a covert channel between an outside attacker and a malware installed on one of the computers in the organization. We showed that even physical obstacles (e.g., high distance, high floor, and hidden scanner) do not prevent an attacker from establishing the described covert channel. We hope our study will increase the awareness to the threat and will result in secured protocols for scanning that will prevent an attacker from establishing such a covert channel specifically in the era of Internet of Things. 
\footnotesize 
\Urlmuskip=0mu plus 1mu\relax
\bibliographystyle{acm}
\bibliography{main}
\end{document}